# Quantum mechanics and free will: counter-arguments

Martín López-Corredoira

Astronomisches Institut der Universitaet Basel, Venusstrasse 7, CH-4102-Binningen, Switzerland
E-mail: martinlc@astro.unibas.ch

**Abstract**

Since quantum mechanics (QM) was formulated, many voices have claimed this to be the basis of free will in the human beings. Basically, they argue that free will is possible because there is an ontological indeterminism in the natural laws, and that the mind is responsible for the wave function collapse of matter, which leads to a choice among the different possibilities for the body. However, I defend the opposite thesis, that free will cannot be defended in terms of QM. First, because indeterminism does not imply free will, it is merely a necessary condition but not enough to defend it. Second, because all considerations about an autonomous mind sending orders to the body is against our scientific knowledge about human beings; in particular, neither neurological nor evolutionary theory can admit dualism. The quantum theory of measurement can be interpreted without the intervention of human minds, but other fields of science cannot contemplate the mentalist scenario, so it is concluded that QM has nothing to say about the mind or free will, and its scientific explanation is more related to biology than to physics. A fatalistic or materialist view, which denies the possibility of a free will, makes much more sense in scientific terms.

## 1. Definition of free will

First of all, we must clarify the meaning of the term to which we refer:

**Free will:** a source totally detached from matter (detached from nature) which is the origin (cause) of options, thoughts, feelings,... That is, the absence of (natural) laws, the existence of an "autonomous mind", i.e. a *principium individuationis*.

This definition is typical of the idealistic tradition, which relates free will with acausality (e.g., Kant 1788). Free will in human beings is opposed to the idea of man as a machine typical of French materialism, for instance in *L'Homme Machine* by La Mettrie (1749).

Other topics which relate to the word "freedom" are not treated here: I do not treat the question of doing what one wants to do. This is a very simple question and has a very simple solution: of course you are free in that sense unless something or somebody forbids you from doing what you want (for instance, if you are in jail). My question is deeper than that: I wonder whether one really wants what he wants, whether the origin of what I want is mine or is an effect of natural laws; whether there is an "ego" separate from nature (dualism). This is a less trivial question and this is the topic which is referred by many classical philosophers when they debate about freedom; Hobbes (1654) and Schopenhauer (1841) are two noteworthy examples.

Quantum mechanics is not metaphysics and we might think that the question of freedom of will is beyond scientific analysis. I agree partially this statement and, as a matter of fact, the aim of this paper is not to engage in metaphysical speculation but the opposite: to refute speculative ideas that are often present in science, particularly in quantum mechanics. I agree with Richard Feynman, the famous physicist, when he says: "If we have an atom that is in an excited state and so is going to emit a photon, we cannot say when it will emit the photon. It has a certain amplitude to emit the photon at any time, and we can predict only a probability for emission; we cannot predict the future exactly. This has given rise to all kinds of nonsense and questions on the meanings of freedom of will, and of the idea that the world is uncertain" (Feynman et al. 1965). Indeed, this paper is not a proposal of using science to solve a metaphysical question once and for all; its purpose is just to state the present–day scientific position concerning freedom of will. Of course, there are other philosophical analyses besides the interpretation of scientific results, but these are not the subject of this paper.

## 2. Indeterminism in quantum mechanics

The discussion about freedom of will usually starts by considering the possibility of an ontological indeterminism. Classical physics is a deterministic model of the world. We can talk about unpredictability but not about a fundamental indeterminism in nature´s laws. However, the orthodox interpretation of quantum mechanics accepts indeterminism in observables once they are measured. For example, we do not know when an atomic nucleus is going to decay. With regard to the wave function, $\psi$, in the interval between two measurements, the evolution is deterministic. The Schrödinger equation is of first order with respect to time. Therefore, from an initial state $\psi(r,t_0)$ in a fixed time $t_0$, we can determine $\psi(r,t)$ for any position $r$ and any time $t$. The indeterminism is not present in the interval between two measurements, only when the measurement is carried out (see, for instance, Cohen–Tannoudji et al. 1977). There are other interpretations which do not need indeterminism in the formulation, such as that by David Bohm (1952), but the most extended interpretation on quantum mechanics accepts an ontological indeterminism in its formulation.

This indeterminism is present in all microscopic systems where quantum effects are important and is diluted in large collections of particles (the macroscopic state) to converge towards Newtonian mechanics, unless there exists a mechanism in which the state of a macroscopic system depends on that of a very few "microscopic" particles.

How can a macroscopic system depend on the state of very few molecules? The answer to this question can be obtained if one applies quantum mechanics to the biology of the human being or any other animal, particularly in the study of the quantum systems in neurology. Ralph Lillie (1927) pointed out possible implications of the quantum indeterminism in macroscopic biological systems which make them different from macroscopic systems with macroscopic components (e.g. the cogwheels of a clock) where the indeterminism is not transmitted on each scale of the system.

Some mechanims have been proposed to make possible an indeterminism in a brain and, therefore, and indeterminism in the whole animal body since the nervous system controls the movements of the muscles in the body. In particular, these mechanisms act in the synapses, in the transmission of neurotransmitters among the neurons of the brain:

- A presynaptic membrane in the terminations of the axons of the neurons controls the triggering of neurotransmitters. These membranes are two molecules thick and have the function of master commuter (Scott 1985) . The small size of the system makes quantum effects important and, hence, indeterminism is present in the membrane as well as in the neurotransmission. Sir John Eccles (1973, 1975, 1994) and Beck & Eccles (1992) back up this theory.

- The microtubules, protein molecules in the dendrites and the axons, are composed of aggregates of particles which can have two different states depending only on the position of one electron. The microtubules take part in the control of the synapse and, therefore, the state of the brain depends on the state of these molecules (Mitchison & Kischner 1984a, 1984b; Penrose 1994; Rosu 1997).

## 3. Mind interacting with the body

Since determinism and denial of free will have traditionally been claimed to be one and the same, some authors thought that quantum indeterminacy would provide a new point of entry for freedom of will that was not allowed by classical physics (e.g. Eddington 1932; Jordan 1944, 1955; Frank 1957; Margenau 1961; Stapp 1995).

If indeterminism affects the behaviour of macroscopic biological systems, this indeterminism will be applicable to the human behaviour and, therefore, Pascual Jordan concluded:

"If the supposition is correct that the controlling reactions of organisms are of atomic physical fineness, it is evident according to our modern knowledge that the organism is quite different from a machine and that its living reactions possess an element of fundamental incalculability and unpredictability. One can object that our fundamental understanding of life phenomena is not greatly aided by considering a statistically functioning dice cup instead of a machine as the pattern of organism. But at the moment it is only important for us to determine in the negative sense that the machine theory of organisms (including their further results; e.g., in the sense of a denial of the freedom of the will) can hardly exist in view of the new physics" (Jordan 1944).

"We can now know the behavior of an individual organism––regardless of whether animal or man––is not exclusively determined by mechanical necessity; we can no longer, with LaMettrie, forbid the soul or the will to intervene in the fixed and predetermined movements of the body's atoms" (Jordan 1955).

Among the different interpretations of quantum mechanics, those that are based on the collapse of the wave function by the mind when this participates in a measurement give rise in some cases to a defense of freedom of will. This idea was proposed by Compton (1935, 1981), von Neumann (1932) and Wigner (1961, 1967) and other authors such as H. P. Stapp (1991, 1993, 1995), L. Bass (1975), W. Heitler (1963), P. J. Marcer (1992), R. Penrose (1994).

Indeed, this is not a totally new idea, this defense of freedom of will resembles extraordinarily the ideas of the Epicureans (for instance, Lucretius' *De rerum natura*), who speculate that the atoms of the body can change their paths according to the will of the mind corresponding to that body. Indeterminism allows the possibility of different paths and the will chooses among them. In the contemporary quantum version, the human will governs the body by means of the presynaptic membranes or the microtubules in the neurons. The synaptic connections are controlled by the mind and the system of neurons is controlled by the synaptic connections (Penrose 1994; Horgan 1994; Rosu 1997).

There are two aspects of the hypothesis of the mind interacting with the body:

- The apparatus cannot produce the collapse of the wave function in the measurement; it becomes entangled (in a superposition of states) with the system being measured. Hence, another kind of element is necessary to produce the collapse of the wave function in the system measurement: namely, the "mind" (e.g. von Neumann 1932; Wigner 1961, 1967). Therefore, the mind can choose the state of all the systems which it is observing. This includes neurotransmission in the brain, which is associated with the mind.

- All the neurons states are coordinated by the mind when it produces a thought. This is argued by a quantum coherence property in the brain, non–local effects on all particles in the brain that confer on it a unity within a unique quantum state (e.g. Stapp 1991, Josephson & Pallikari–Viras 1991).

This is an interactive dualism, as in Descartes (1641) or Popper & Eccles (1977). The presynaptic membranes, or microtubules, play a role similar to that of Descartes' point of connection between mind and body. In Popper & Eccles, we find indeed a new version of the hypothesis of mind interacting with the body, but they claim that the interaction is located in a particular region of the left hemisphere of the brain instead of all the neurons in the brain. In any case, the characteristics of the quantum subjectivists, Descartes or Popper & Eccles are all similar: they are interactive dualists. Some authors who are in favour of free will on the basis of quantum mechanics deny being "dualists" (e.g. Wigner) and prefer to be called "emergentists", but their emergentism is in some way concealed dualism.

## 4. Counter–arguments (I): indeterminism does not imply free will

There are two options: determinism or indeterminism. In the case of determinism, the atoms of our body follow strictly deterministic physical laws, and there is no possibility of our intervening; we cannot generate the first cause for the motion of the atoms in our body, we are locked in a Laplacian mechanicism that does not leave space for freedom of will (for freedom of will as defined in the first section, not for naive popular notions about freedom).

The case we are interested now is the second one, that of indeterminism. Let us imagine that quantum mechanics gives a correct theoretical framework in which to defend ontological indeterminism (not merely unpredictability). What about freedom of will in this case? Here, there might be free will but not necessarily because "indetermism does not imply free will." Of course, indeterminism is a necessary condition for freedom of will but it is not sufficient. The incompatibilist tradition long thought about the determinism when the question of free will arose and argued against determinism; this led many people to believe that determinism is the opposite of freedom of will. Among those authors confused by these implications were those who quickly applauded the achievements of quantum mechanics because it opened a new door on free will.

Assuming that discussion of free will is necessarily a discussion concerning necessity or contingency is wrong. As Kant said, freedom is neither nature nor chance. Philosophers or scientists who think that indeterminism gives freedom of will forget the rules of classical logic and claim that "(p --> ¬q) implies (¬p --> q)", where "p" stands for determinism and "q" stands for freedom of will. This argument is false.

For example, let us imagine building a robot that follows random laws. Is it free? Of course not. Indeterminism is not an absence of causation but the presence of non-deterministic causal processes (Fetzer 1988). I mean that "causality" is not necessarily determinism; we can understand "causality" in a more general sense: causality as "explanation" or "reason". An explanation of or the reason for an event means following a law (perhaps a statistical law), and the presence of laws is the absence of free will. Quantum mechanics is indeterministic but it is not acausal. There is always a cause, an explanation or reason, for any phenomenon; for example, when an electron which is pushed towards another electron. Both electrons are repelled, and their positions and velocities are undetermined. The cause of repulsion is that we joint both electrons. The electrons are not free to choose their repulsion.

Giving up fatalism derived from scientific materialism requires avoiding any idea of causality, avoiding any possible explanation for phenomena. When an act or election can be explained in terms of physical laws (even probabilistic laws), then we are including this action or election as a natural phenomenon. Therefore, we deny that the origin of this action or election comes from ourselves as something independent of nature, i.e. we deny free will.

## 5. Counter-arguments (II): scientific knowledge about consciousness

Any argumentation in favor of an autonomous mind based on modern physics is gratuitous; i.e. it is mere opinion rather than an empirically established result.

Wigner, von Neumann and many others have pointed out that consciousness is necessary to understanding quantum mechanics in order to avoid contradictions, but their argumentation is somewhat dubious. The differentiation between systems that entangle mutually and systems that produce wave function collapse when coupled stands to reason in the formalism of quantum mechanics. I also agree that human beings are present in a final state of the measurement, when the results are checked. Nevertheless, I reject the statement that mind is the agent of wave function collapse. Measurement is associated with collapse and human beings are present in the final stage of the measurements, but this does not imply that human beings produce collapse.

What else can produce collapse? Is there any alternative to the hypothesis of mind producing collapse? Yes, there is. We can talk about collapse without making claims concerning the intervention of the consciousness (Stenger 1997). A materialist and reductionist position is fully consistent with the observed facts and quantum theory. A human observer in the measurement is not even necessary (e.g. Shimony 1988, Mulhauser 1995). There are alternatives; for instance, the measuring apparatus coupled with the measured system can be collapsed with the wave function in the microscopic components when the latter is measured in a macroscopic domain. Indeed, Bohr´s interpretation, as well as that of most of present-day leading specialists in quantum physics, is that the central element in the measurement is not the consciousness but the distinction between the measured system and the measuring apparatus. A computer could produce the measurement and the observer could read the results once they were already obtained and saved in the memory of the computer. This cannot be checked, since we will never know whether the collapse was produced by the computer and the measuring apparatus before the observer checked the results or by the human mind when the observer checked the results (Schrödinger´s cat paradox), but at least we know that we can interpret quantum mechanics without the notion of an autonomous mind in the role of an observer.

We should avoid confusion between coherence in a quantum system, in which the different parts are entangled and mutually dependent until collapse is produced, with the idea of a "conductor", which governs the system, as in the case of a conductor of an orchestra. The parts of a physical system could be compared to the musicians of an orchestra, in which everybody listens to everybody, but the cohesion would be noticeable only for very low energies (low temperature) or when the interaction with the surroundings is very faint (Mulhauser 1995). Even if this was the case in a human brain, such as superconductors with low temperature, we have no conductor. Neither classical nor quantum physics has anything to say about a hypothetical conductor (mind) governing the brain. Subjectivist arguments about the mind/body problem by H. P. Stapp (1991) based on the difference between classical and quantum physics cannot sustain an argument in favour of freedom of will.

We must conclude that contemporary physics has not succeeded in approximating further to a knowledge of an autonomous consciousness that freely governs the body. The introduction of the new notions of quantum mechanics is irrelevant to the mind/body problem (Mulhauser 1995, Ludwig 1995). You could adopt a subjectivist position, just as somebody could be a defender of Berkeley´s idealism in the XVIII century, but these are simply opinions beyond scientific debate.

Physicists have nothing to say about the mind but neurologist do. Nowadays, most neurologists insist that the idea of a soul or an autonomous mind is a myth. They adopt a materialistic philosophy in which the mind can be explained in terms of neurological processes (Crick 1994).

There are plenty of examples to illustrate the materialistic philosophy in the neurological sciences. Electrical signals applied to the brain produce variations in the conciousness such as images and recollections (e.g., Penfield & Perot 1963). The opposite is also true: electric signals are registered in the brain for any conscious sensation. Indeed, electroencephalograms register activity before the mind is aware of the sensation. For instance, in an experiment (Deeke et al. 1976; Libet 1985, 1987) in which some volunteers connected to electroencephalograms were asked to move their fingers at will, the apparatus registered electrical activity around 300 milliseconds before the person was conscious of taking a decision to move a finger. This clearly implies that unconscious activities of the brain trigger activity of the neurons. And this means that matter governs conciousness before conciousness governs matter. The mind cannot be autonomous in the decisions. Moreover, it is clear that there is a minimum time (around 60 or 70 milliseconds) for the brain to compute the most simple sensation (Libet 1985). This time interval is too great to be interpreted as a spontaneous decision. The neurons cannot wait such a long time for an order from the mind to choose a state in the neurotransmission.

The property of a quantum coherence state in the whole brain is also absurd if we think about experiments where the junction between both hemispheres of the brain is cut. In this case, two wills are produced (Sperry 1964). The unity of the brain is produced by the "neurological connections" rather than by a property of quantum coherence. Neither can we talk about the relation of the will (the capacity for taking decisions) to the capacity of observing in a measurement, as quantum subjectivity proposed. There are areas in the brain (the 24th Brodmann area) related directly to the will and when these areas are affected the patient can observe ("measure") but the brain cannot send any order to the body to be moved (Crick 1994, post scriptum).

All this is experimental evidence, not just hypothesis or speculation. And all these facts point out that there is no autonomy in the brain, there is no "ego" sending orders to the body. The defense of free will is impossible in this context, and speculation about a subjectivist role of consciousness in measurement is absurd.

## 6. Counter–arguments (III): Evolution and ontogeny

The problem of most of the philosophers who defend an origin of the mind in the context of evolutionary theory is that they do not know this theory or they forget many of its important points. Most often, they do not worry about the origin of the mind. It is there–they think–and has "emerged" from matter in some way; they think that the word "emergence" solves all the problems that leave them free to talk about the "spiritualization" of matter. The same question arises when we wonder about the origin of the mind during the growing of a human body through cell multiplication (ontogeny) from the moment that the spermatozoid and the ovule join. If we forget religious and personal beliefs, we must consider that evolution and cell multiplication cannot explain the emergence of an autonomous mind collapsing matter from DNA copying, which just sequences the building of proteins in the different tissues, and mutations. It is also futile to repeat the word "emergence" insistently while we are in the world of "matter". The emergence of what? Of a mind that can be distinguished from matter but at the same is generated by the matter? Absurd!

According to the quantum subjectivism, the mind should emerge instantaneously. There is no place for half or a quarter of a mind producing half or a quarter of a collapse. Either mind produces collapse, or there is no mind and matter can only become entangled with other physical systems. This is the position of von Neumann, Wigner, etc. Then, we must think about a spontaneous creation of the mind. We have a baby without mind, it is just a piece of matter. One second later, we have a baby with a mind that can produce the collapse of the wave function in the systems which he observes. Absurd!

The most difficult question to solve is the paradox of the Universe before the existence of any mind (Bohm & Hiley 1993). If the mind produced the collapse of wave functions in matter, then nature before the existence of minds was uncollapsed, and there was no birth of the Universe because it is still in a superposition of states. Absurd! Further ridiculous ideas were proposed to explain this paradox (e.g., Kafatos & Nadeau 1990) by arguing that some Universal Mind (God?) was present before the existence of life on the earth to collapse the wave functions, but this pantheist solution does not explain why human mind is now responsible of the collapse instead of God´s Mind. Did He take a holiday after our appearance? Absurd!

## 7. Conclusions

- Indeterminism does not imply free will.

- The opposite of free will is materialism rather than determinism.

- Dualism and "mind collapsing matter" from quantum subjectivism is against observational evidence in neurology.

- Dualism and "mind collapsing matter" from quantum subjectivism is against evolution theory.

- The contemporary scientific position no more has a place for freedom of will than French materialism of XVIII century.

**Acknowledgements:**

This article has been revised for english and style by Terry Mahoney (Instituto de Astrofísica de Canarias).